# Machine learning with force-field inspired descriptors for materials: fast screening and mapping energy landscape

Kamal Choudhary, Brian DeCost, Francesca Tavazza

1 Materials Science and Engineering Division, National Institute of Standards and Technology, Gaithersburg, Maryland 20899, USA

## ABSTRACT

We present a complete set of chemo-structural descriptors to significantly extend the applicability of machine-learning (ML) in material screening and mapping energy landscape for multicomponent systems. These new descriptors allow differentiating between structural prototypes, which is not possible using the commonly used chemical-only descriptors. Specifically, we demonstrate that the combination of pairwise radial, nearest neighbor, bond-angle, dihedral-angle and core-charge distributions plays an important role in predicting formation energies, bandgaps, static refractive indices, magnetic properties, and modulus of elasticity for three-dimensional (3D) materials as well as exfoliation energies of two-dimensional (2D) layered materials. The training data consists of 24549 bulk and 616 monolayer materials taken from JARVIS-DFT database. We obtained very accurate ML models using gradient boosting algorithm. Then we use the trained models to discover exfoliable 2D-layered materials satisfying specific property requirements. Additionally, we integrate our formation energy ML model with a genetic algorithm for structure search to verify if the ML model reproduces the DFT convex hull. This verification establishes a more stringent evaluation metric for the ML model than what commonly used in data sciences. Our learnt model is publicly available on the JARVIS-ML website (https://www.ctcms.nist.gov/jarvisml ) property predictions of generalized materials.

## I. INTRODUCTION

Machine learning has shown a great potential for rapid screening and discovery of materials [1]. Application of machine learning methods to predict material properties has started to gain importance in the last few years, especially due to the emergence of publicly available databases [2-6] and easily applied ML algorithms [7-9]. Chemical descriptors based on elemental properties (for instance, the average of electronegativity and ionization potentials in a compound) have been successfully applied for various computational discoveries such as alloy-formation [10]. Nevertheless, this approach is not suitable for modeling different structure-prototypes with the same composition because they ignore structural information. Structural features have been recently proposed based on Coulomb matrix [11], partial radial distribution function [12], Voronoi tessellation [13], Fourier-series [14], graph convolution networks [15] and several other recent works [16-20]. However, none of these representations explicitly include information such as bond-angles and dihedral angles, which have been proven to be very important during traditional computational methods such as classical force-fields (FFs) [21] at least for the extended solids. Hence, we introduced those descriptors in our ML model. Additionally, we are also introducing charge-based descriptors, inspired by classical-force field community such as charge-optimized many-body potentials (COMB) [22], reaction-force fields (ReaxFF) [23] and Assisted Model Building with Energy Refinement (AMBER) [24]. We first introduce a unique set of classical force-field inspired descriptors (CFID). Then, we give a brief overview of gradient boosting decision tree algorithm (GBDT) and JARVIS-DFT database on which CFID is applied. After that, we train two classification and twelve regression models for materials properties. We use the regression models to screen new 2D-layered materials based on chemical complexity, energetics and bandgap. We verify the machine learning predictions with actual density functional theory calculations. Finally, we integrate a genetic algorithm with formation energy machine learning model to generate all possible structures of few selected systems. The energy landscape in terms of convex hull plot from the machine learning model is in great agreement with that from actually density functional theory calculations. This leads to a new computationally less expensive way to map energy landscape for multicomponent systems.

## II. CLASSICAL FORCE-FIELD INSPIRED DESCRIPTORS (CFID)

We focus on development of structural descriptors such as radial distribution function, nearest neighbor distribution, angle and dihedral distributions, and we combine them with chemical descriptors, such as averages of chemical properties of constituent elements and average of atomic radial charge (like COMB/ReaxFF formalisms), to produce a complete set of generalized classical force-field inspired descriptors (CFID). The radial distribution function (RDF) and neighbor distribution function are calculated for each material up to 10Å distance. Bond-angle distributions (ADF) are calculated for "global" nearest neighbors (ADF-a) and for "absolute" second neighbor (ADF-b). For multi-component systems, we define as "global nearest neighbor distance" as the distance that includes at least one pair interaction for each combination of the species (AA, AB, and BB for an AB system, for instance). Conversely, the "absolute second neighbor distance" only includes the first two shells of neighbors, irrespective of their specie-type. Dihedral angle distributions (DDF) are included to capture four-body effects and are only calculated for the global first neighbors. We assume that the interatomic interactions are important only up to four-body terms, and higher order contributions are negligibly small. For every single element, we obtained the atomic radial charge distribution from 0 to 10 Å from the pseudopotential library [25]. The average of the charge distributions for all constituent elements in a system gives a fixed-length descriptor for the material. A pictorial representation of the CFID descriptors used here is given in

Fig. 1. A full list of chemical features is given in Table S1 [26]. We also take the sum, difference, product, and quotient of these properties leading to additional chemical descriptors. We cover 82 elements in the periodic table for chemical descriptors. The total number of descriptors found by combining the structural and chemical descriptors is 1557. It is to be noted that the CFID is independent of using primitive, conventional or supercell structures of a material, hence, it provides great advantage over many conventional methods such as Coulomb matrix where primitive structure must be used for representing a material [27].

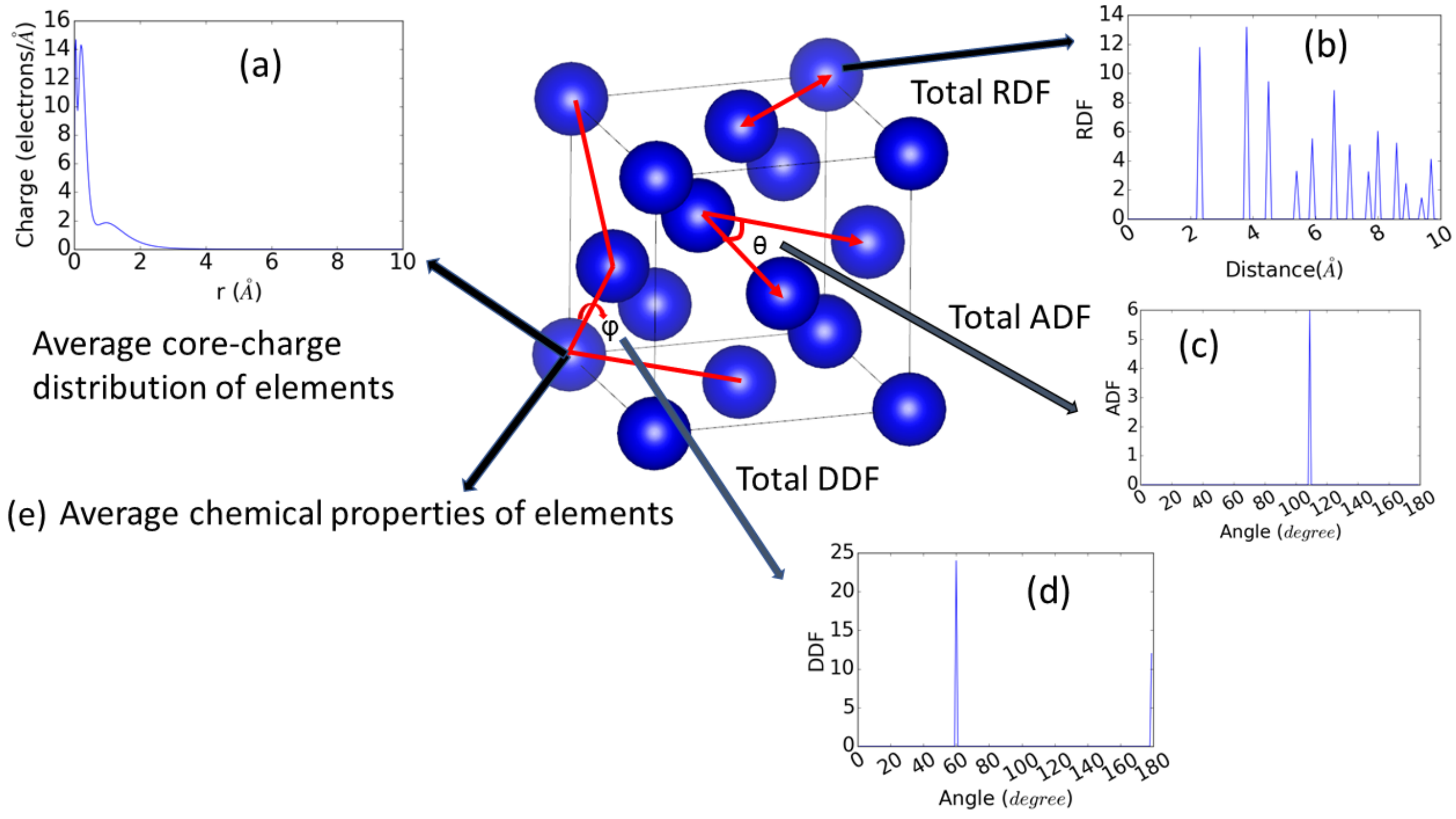

*Fig. 1 Figure showing different components of classical force-field inspired descriptors (CFID) for Si diamond structure. a) average radial-charge density distribution of constituent elements. b) total radial distribution function of the crystal structure, c) total angle distribution function up to*

*the first-nearest neighbor, d) total dihedral-angle distribution up to the first-nearest neighbor, e) average chemical properties of constituent elements. The nearest neighbor distribution was obtained like the RDF.*

## III. TRAINING DATA AND ALGORITHM

For model training, we use our publicly available JARVIS-DFT database [5] which (at the time of writing) consists of 24549 bulk and 616 monolayer 2D materials with 24549 formation energies, 22404 OptB88vdW (OPT) and 10499 TBmBJ (MBJ) bandgaps and static dielectric constants [28], 10954 bulk and shear modulus [29] and 616 exfoliation energies for 2D layered materials. The database consists of multi-species materials up to 6 components, 201 space groups, and 7 crystal systems. Moreover, the dataset covers 1.5 % unary, 26% binary, 56 % ternary, 13 % quaternary, 2 % quinary and 1% senary compounds. The number of atoms in the simulation cell ranges from 1 to 96. To visualize the descriptor data, we perform- t-distributed stochastic neighbor embedding (t-SNE) [30]. The t-SNE reveals local structure in high dimensional data, placing points in the low-dimensional visualization close to each other with high probability if they have similar high-dimensional feature vectors. Results obtained with complete CFID descriptors for all the materials in our dataset are shown in Fig. 2a; the marker colors indicate the crystal system of each material. These plots clearly demonstrate that our database is well-dispersed, and that the data are not biased in favor of a particular type of material. Additionally, materials with similar chemical descriptors tend to be correlated in terms of crystal structure as well. We also visualize the range of target property data. An example of formation energy is shown in Fig. 2b. Clearly, the data is more centered around -4 to 2 eV/atom. Target property distributions of other properties are given in the supplementary information (Fig. S1 [26]).

Of the many ML algorithms available to date, only a small fraction offers high interpretability. To enhance interpretability of the ML models we chose gradient boosting decision trees (GBDT) method [25]. The GBDT method allows obtaining the feature importance for training which can be used to interpret the guiding physics of a model. In this work, we use two classifications and twelve independent regression models with gradient boosting decision tree (GBDT) [9,31,32]. The GBDT model takes the form of an ensemble of weak decision tree models. Unlike common ensemble techniques such as AdaBoost and random forests [32], the gradient boosting learning procedure consecutively fits new models to provide a more accurate estimate of the response variables. The principal idea behind this algorithm is to build the new base learners to be maximally correlated with the negative gradient of the loss function, associated with the whole ensemble. Suppose there are N training examples: $\{(x_i, y_i)\}^N$ then GBDT model estimates the function of future variable x by the linear combination of the individual decision trees using:

$$f_m(x) = \sum_{m=1}^{M} T(x; \theta_m) \tag{1}$$

Where $T(x; \theta_m)$ is the i-th decision tree, $\theta_m$is its parameters and M is the number of decision trees.

The GBDT algorithm calculates the final estimation in a forward stage-wise fashion. Suppose the initial model of x is $f_0(x)$, then the model in m step can be obtained by the following relation:

$$f_m(x) = f_{m-1}(x) + T(x; \theta_m) \tag{2}$$

where $f_{m-1}(x)$ is the model in (m−1) step. The parameter $\theta_m$ is learned by the principle of

empirical risk minimization using:

$$\widehat{\theta_m} = \arg \min_{\theta_m} \sum_{i=1}^{N} L\big(y_i, f_{m-1}(x) + T(x; \theta_m)\big) \tag{3}$$

where L is the loss-function. Because of the assumption of linear additivity of the base function, we estimate the $\theta_m$ for best fitting the residual $L\big(y_i, f_{m-1}(x)\big)$.

The parameters of a decision tree model are used to partition the space of input variables into homogeneous rectangular areas by a tree-based rule system. Each tree split corresponds to an if-then rule over some input variables. This structure of a decision tree naturally models the interactions between predictor variables. At each stage, parameters are chosen to minimize loss function of the previous model using steepest descent. As a standard practice, we use train-test split (90 %:10 %) [33,34] , five-fold cross-validation [10] and examining learning curve (Fig. S2 [26]) in applying the GBDT with CFID. The 10 % independent test set is never used in the hyperparameter optimization or model training so that the model can be evaluated on them. We performed five-fold cross-validation on the 90 % training set to select model hyperparameters. During training, we use the early stopping regularization technique to choose the number of decision trees ($T(x;\theta_m)$ ): we grow the GBDT model by 10 trees at a time until the mean absolute error (MAE) on the validation set converges. Then other hyperparameters such as learning rate and the number of leaves of GBDT are optimized via the random search of five-fold cross-validation with the optimal number of trees from the previous step. The optimized model is used to produce learning curve of the model to check if the model can improve by addition of data. Finally, the feature importance of all the descriptors is obtained with GBDT to interpret the importance of various descriptors in training a model. Additionally, we provide comparison of learning curves for OPT and MBJ bandgap learning curves in Fig. S3 [32]. We observe that for similar data-sizes, the MBJ bandgap ML model still has higher MAEs than OPT ML model. The learning curves in Fig. S2 [35] can be used to examine training size dependent accuracies of various models.

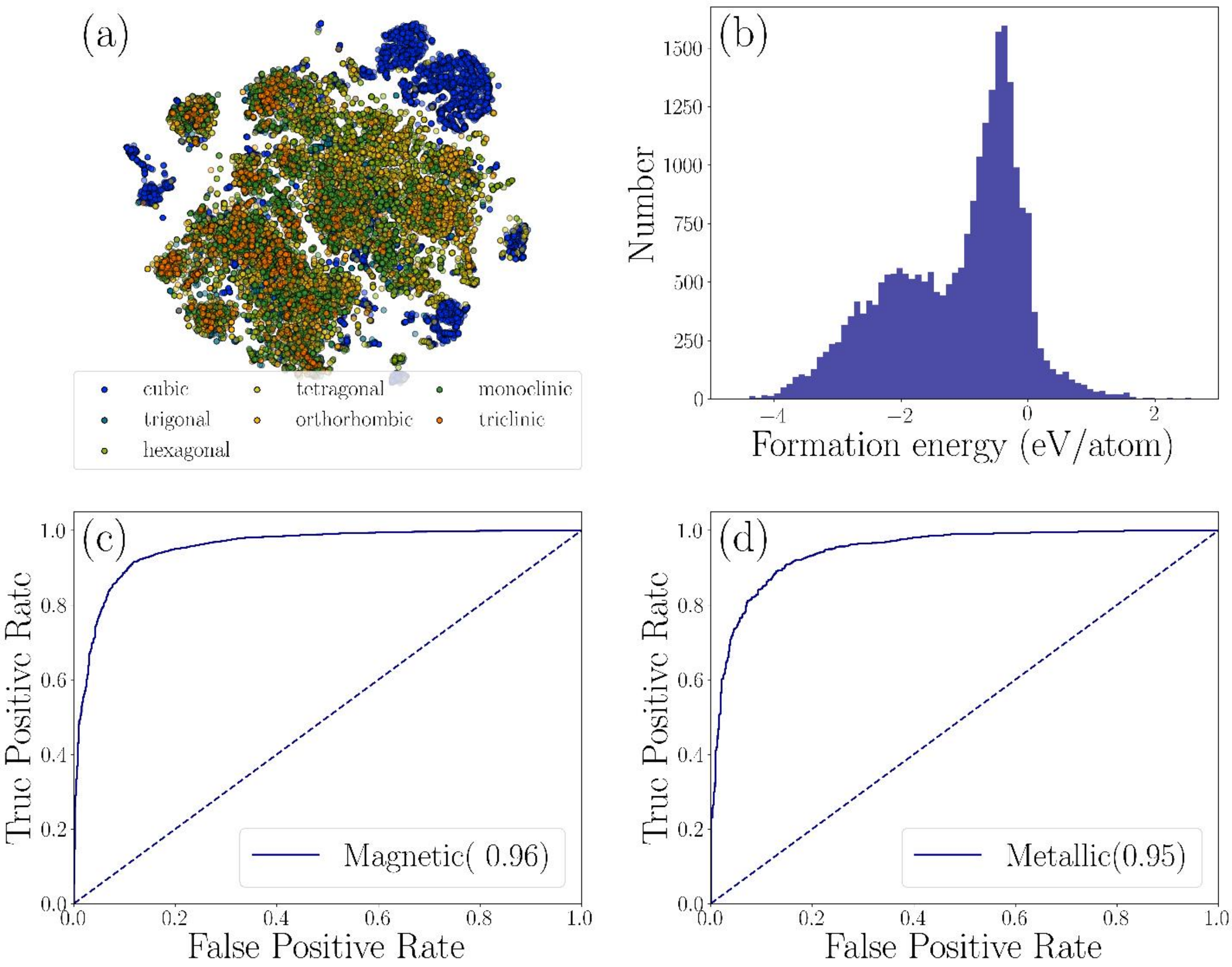

*Fig. 2 Visualization of data and classification problems. a) t-SNE plot, b) histograms for formation energy distribution, c) ROC-curve for metal/insulator classification and d) ROC-curve for magnetic/non-magnetic material classification.*

## IV. MODEL PERFORMANCE AND INTERPRETATIONS

To apply the CFID descriptors, we tested metal/insulator and magnetic/non-magnetic classification problems. The performance of the classification model is measured from the area under the receiver operating characteristic (ROC) curve. For metal/insulator and magnetic/non-magnetic classification problems, we obtained the area as 0.95 and 0.96, respectively (fig. 2c and 2d). The results clearly show that the successful applications of CFID for material classifications.

In addition to predicting exact bandgap ($E_g$) values (using regression) and then screen materials, we can simply classify materials into metallic ($E_g$=0) and non-metals ($E_g$>0). Similar classification can be applied for magnetic/non-magnetic systems.

*Table. 1: Statistical summary of different regression models. We report the number of data points and mean absolute error (MAE) of classical force-field inspired descriptor (CFID) models on 10% held data, MAE of DFT predictions compared to experiments, mean absolute deviation (MAD) of the test data (DFT). The bandgaps and refractive indices were obtained with OptB88vdW (OPT) and Tran-Blaha modified Becke-Johnson potential (MBJ). All other quantities were obtained with OPT only.*

| **Property** | **#Data-points** | **$MAE_{CFID-DFT}$** | **$MAE_{DFT-Exp}$** | **$MAD_{DFT}$** |
|---|---|---|---|---|
| **Formation energy (eV/atom)** | 24549 | 0.12 | 0.136 [13] | 0.809 |
| **Exfoliation energy (meV/atom)** | 616 | 37.3 | - | 46.09 |
| **OPT-bandgap (eV)** | 22404 | 0.32 | 1.33 [28] | 1.046 |
| **MBJ-bandgap (eV)** | 10499 | 0.44 | 0.51 [28] | 1.603 |
| **Bulk modulus (GPa)** | 10954 | 10.5 | 8.5-10.0 [29,36] | 49.95 |
| **Shear modulus (GPa)** | 10954 | 9.5 | 10.0 [29,36] | 23.26 |
| **OPT-$n_x$ (no unit)** | 12299 | 0.54 | 1.78 [28] | 1.152 |
| **OPT-$n_y$ (no unit)** | 12299 | 0.55 | - | 1.207 |
| **OPT-$n_z$ (no unit)** | 12299 | 0.55 | - | 1.099 |
| **MBJ-$n_x$ (no unit)** | 6628 | 0.45 | 1.6 [28] | 1.025 |
| **MBJ-$n_y$ (no unit)** | 6628 | 0.50 | - | 0.963 |
| **MBJ-$n_z$ (no unit)** | 6628 | 0.46 | - | 0.973 |

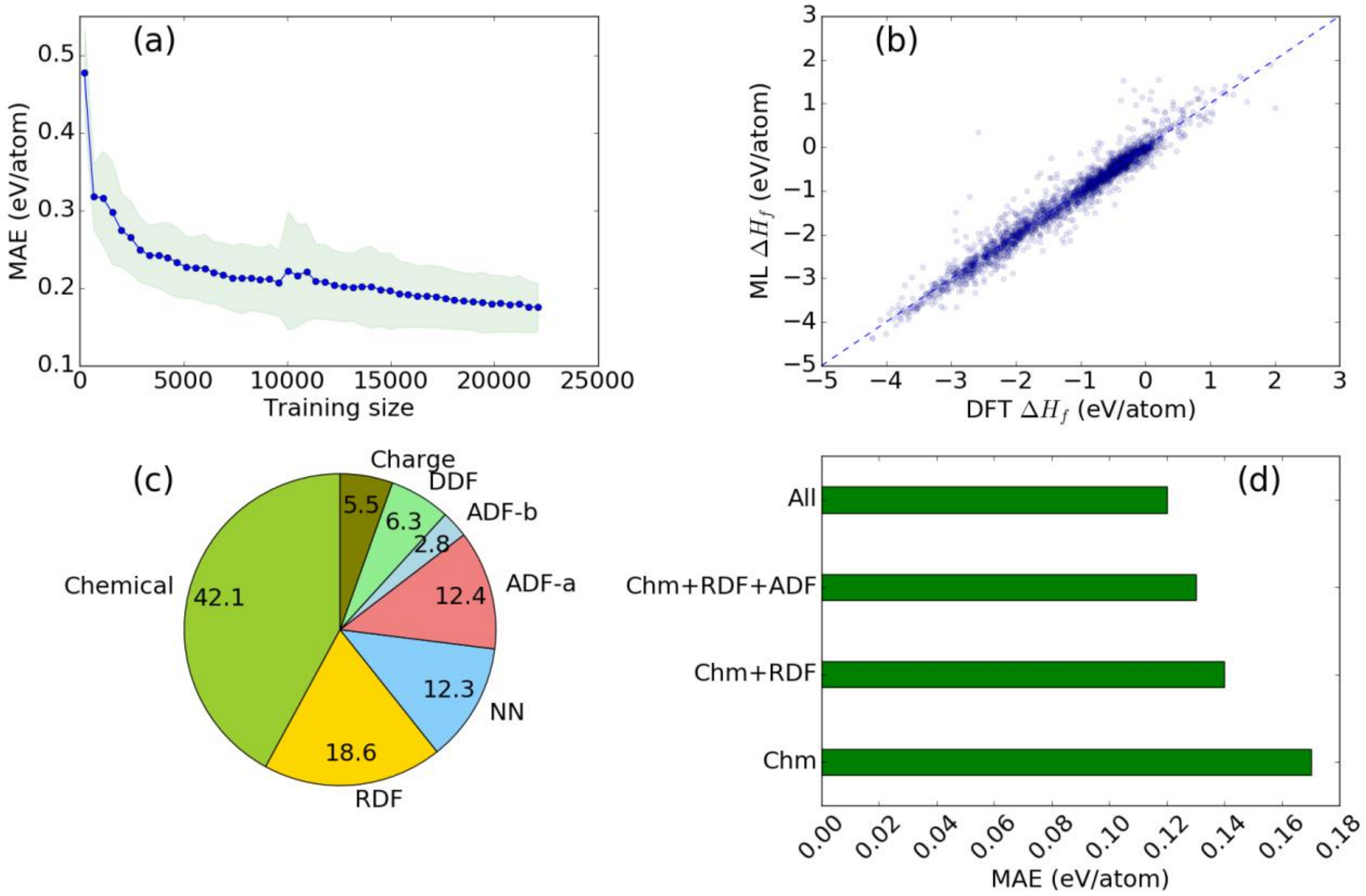

*Fig. 3 Performance and interpretation of formation energy ML model. a) learning curve, b) ML prediction on 10 % held samples, c) group wise feature importance of descriptors, d) comparison of model performance by incrementally adding various structural descriptors.*

Next, we perform twelve independent regression tasks on above-mentioned properties. The mean absolute error (MAE) results obtained from applying the models on 10 % held set) are shown in

Table. 1. Because each property has different units and in general a different variance, we also report the mean absolute deviation (MAD) for each property to facilitate unbiased comparison of the model performance between different properties. The MAE and MAD values were computed as:

$$MAE = \frac{1}{n}\sum_{i=1}^{n}|x_i - y_i| \tag{4}$$

$$MAD = \frac{1}{n}\sum_{i=1}^{n}|x_i - \bar{x}| \tag{5}$$

$$\bar{x} = \frac{1}{n}\sum_{i=1}^{n} x_i \tag{6}$$

For MAE calculations, $x_i$ represents the predicted ML predicted data and $y_i$ the DFT data for the i-th sample. The MAD calculations ($MAD_{DFT}$) are intended as a robust estimate of the DFT values. While MAE shows the accuracy of the models, the MAD helps understand the statistical variability in the data. Clearly, all the ML model-uncertainties ($MAE_{CFID-DFT}$, $\delta_{ML}$) are comparable to the experimental error of DFT predictions ($MAE_{DFT-Exp}$, $\delta_{DFT}$). We assert that the MAEs obtained here are acceptable for screening purposes. The ML MAE values do not directly compare with DFT, because the reference data for DFT is experimental data while the reference for ML models is the DFT data. However, the MAEs can help identify the range in predicted values: our CFID GBDT model fits the DFT training data about as well as the DFT itself matches experimental data. Also, assuming the error in DFT and ML to be independent, the compound uncertainties can be given as:

$$\delta = \sqrt{(\delta_{ML}^2 + \delta_{DFT}^2)} \tag{7}$$

Currently, there are several formation energy ML models in the literature [13,15] with MAE ($\delta_{ML}$) ranging from 0.039 to 0.25 eV/atom. We assume that the MAE should be independent of different datasets because the structures originate from the ICSD database. The MAE of our model (0.12 eV/atom) is in the same range as all of those, and its learning curve (shown in Fig. 3a) clearly shows that the model can be further improved by adding more data. We have achieved comparable ML model accuracy by incorporating additional domain knowledge (i.e. structural features in addition to chemical features).

Our bandgap model predictions for OPT (0.32 eV) is better than MBJ (0.44 eV) mainly because of the number of data points included during training (19782 for OPT versus 9546 for MBJ). In both cases, metals and non-metals were included during training. In general, MBJ ML model should be preferred to predict band gaps because of the inherent band gap underestimation problem in OPT [28]. However, the MAE of this model is slightly larger than the OPT one right now because its training set is almost half. As we add more data, we expect to decrease the MAE.

We also demonstrate the applicability of ML models for predicting static refractive indices and exfoliation energies. The OPT and MBJ refractive index models were trained for non-metallic systems only because DFT methods generally do not consider intra-band optoelectronic transitions. To our knowledge, we are the first to apply ML to predictions of refractive indices and exfoliation energies. Our MAE for the refractive indices is between 0.45 to 0.55, depending on the model (OPT or MBJ) and crystallographic direction. We monitor the MAE during the learning curves as they reach a plateau. Interestingly, we achieved a very accurate refractive index model (reaching the plateau) with training sets of the order of $10^3$, while the models for all the other examined quantities required training sets of the order of $10^4$ to achieve high accuracy. However, specific hyperparameter and learning curve dependence on a particular type of target data in a ML model is beyond the scope the present paper. Generally, these axes are well-defined from experiments (X-ray diffraction, ellipsometry and similar techniques), so the average of the refractive indices in x, y and z-crystallographic dimensions can be compared to experimental data. Also, training on individual refractive indices allow to predict anisotropy in optical property data. Our work proves that though having relatively smaller dataset, highly accurate ML models can be obtained with CFID descriptors because of the chemo-structural information. Generally, more the data more accurate the ML models, but we show by adding detailed domain knowledge can also improve accuracy in the materials domain. Additionally, the idea is to screen materials based on several properties such as formation, energy, bandgap, refractive index, exfoliation energy and magnetic moment etc. with fast ML models, which in regular DFT or other methods require separate calculations and hence ML can accelerate the process.

Recently, 4079 materials have been predicted to be layered using the data-mining and lattice constant approach [5,37]. Exfoliation energies are ultimately needed to computationally confirm whether a material is exfoliable or not. A material is considered exfoliable if its exfoliation energy is less than 200 meV/atom. As such DFT calculations are very expensive, we only have 616 DFT calculated exfoliation energies, which makes for a very small training set. Our MAE for exfoliation energy ML model is 37 meV/atom. Given that the threshold for a material to be exfoliable is 200 meV/atom, our MAE is reasonable for initial screening of exfoliable materials. Our bulk and shear modulus models have MAEs that are comparable to DFT MAE (10 GPa) [29,36] and previous ML models (9 GPa and10 GPa) [38]. It is to be noted that 2494 descriptors were used in Isayev et al. [38] model, while a comparable accuracy was achieved here with fewer descriptors.

Next, we interpret our ML models using feature importance analysis for structural, chemical and charge descriptors, as shown in Fig. 3c in the case of formation energies. Not surprisingly, the chemical features are found to be the most important during training. Chemical descriptors such as average of heat of fusion, boiling and melting point of constituent elements along with cell size based descriptors such as packing fraction and density of the simulation cell play very important role in providing accurate models. Although chemical descriptors are the major players in determining the accurate model, RDF and ADF are also found to be very important. Interestingly, the charge descriptors were found to be the least important. Further analysis shows that radial distribution function (6.8 Å bin, 5.5 Å bin), nearest neighbor (5.5 Å bin) angle distribution (178$^o$, 68$^o$) and DDF (43$^o$ and 178$^o$) were found as some of the most important structural features of the formation energy model. This is intuitively tangible because angles such as 60$^o$ and 180$^o$ are key in differentiating materials such as FCC and BCC crystals. The RDF and NN contribution for 0 Å to 0.6 Å play the least important role among all the RDF and NN descriptors. This is also obvious as no bond-length exists at such small distances. We find that number of unfilled d and f orbitals-based descriptors play important roles in classifying magnetic/non-magnetic nature of a material. We have added feature importance of different models to compare their importance in training different models in the supplementary information [26]. We observe that quantities such as formation energy, modulus of elasticity, refractive index are highly dependent on density of the simulation cell, RDF, ADF, packing fraction while quantities such as bandgap, magnetic moment are mainly dependent on chemical property data, as seen by top ten descriptors of each model in SI. Based on the above argument, we claim that our models can capture important physical insights of a problem though they are primarily data-driven.

To quantify the effect of introducing structural descriptors, we train four different formation energy models by incrementally adding structural descriptors: a) average chemical and charge descriptors (Chm) only, b) Chm with RDF and NN, c) Chm with RDF, NN and ADF, and d) including all the descriptors. The MAE of these models is shown in Fig. 3d. We observe that as we add more structural descriptors, the MAE gradually decreases. The lower MAE values clearly establish that there is indeed improvement due to the introduction of structural descriptors. The trained model parameters for each model was saved and can be used to make predictions on arbitrary materials. An interactive web app for predicting the formation energy and properties of arbitrary materials based on the trained CFID GBDT models is available at https://www.ctcms.nist.gov/jarvisml/ . The training data and code for ML training is already available at: https://github.com/usnistgov/jarvis .

## V. SCREENING OF 2D-MATERIALS AND INTEGRATING GENETIC ALGORITHM

As an application, we use the ML models to discover new semiconducting 2D layered materials. We first obtain all the 2D materials predicted from lattice constant approach [5] and the data-mining [37] approaches. This results in 4079 possible candidates. Only a few hundreds of them have been computationally proven to be exfoliable yet because exfoliation energy calculations in DFT are computationally expensive. The above-mentioned approaches can be combined with ML models to screen 2D layered materials. For example, using out trained ML models, we successively screen materials to have MBJ bandgaps in the range of 1.2 eV to 3 eV, then negative formation energies and lastly exfoliation energies less than 200 meV/atom. This procedure quickly narrows down the options to 482. At this point, we chose structures with the number of unique atom-types less than 3 (to lessen complexity in future experimental synthesis), which resulted in 65 candidates. Some of the materials identified by this screening procedure were CuI (JVASP-5164), $Mo_2O_5$ (JVASP-9660) and InS (JVASP-3414). To validate, we calculated exfoliation energy for CuI using DFT (as an example case) on bulk and single layer counterparts and found the exfoliation energy to be 80.0 meV/atom, which confirmed that it should be exfoliable 2D materials. However, we found that for InS and $Mo_2O_5$, the DFT exfoliation energy were 250 and 207 meV/atom, which is not too high from 200 meV/atom cut-off. We have already found several other iodide, oxide and chalcogenide materials using the lattice constant criteria [5]. These examples show that the DFT application, in series, of the ML models for various physical properties can significantly accelerate the search for new materials for technological applications.

Lastly, we feel it is important to point out that, although the accuracy metrics presented in Table 1 are compelling from a data science perspective, the metrics may not be sufficient for materials science applications, as there are many physical constraints that should be satisfied as well. The most important of them is the identification of the correct ground state structure. For instance, a face-centered cubic (FCC) Cu should be ground state among all the possible combinations/rearrangement of Cu atoms. To include this metric in our models, we integrate our ML model with genetic algorithm (GA) search [39,40] to produce a large number of possible structures. In the Cu example, we start with Cu structure prototypes such as FCC, body-centered cubic (BCC) Cu and let it evolve using GA. After 5000 structure-evaluations, we found only one phase of Cu in ML prediction to be more stable than FCC Cu. This phase turns out to be the metastable tetragonal Cu phase (space group I4/mmm) shown in Fig. S4 [26]. The tetragonal structure was also observed during Bain-path study of Cu-system [41]. We carried out DFT calculation on this structure and found that the structure was only 0.01 eV/atom higher in energy than the FCC phase. This energy difference value lies much below than MAE of our ML formation energy model, and therefore validate the applicability of our ML approach. Such a GA-search is not feasible in ML models with only chemical descriptors. We did a similar search for Mo-S system as well. We used the known prototypes of Mo-S systems as parents and produced offspring structures using GA. Our goal was to check if the ML models find the same ground state structure as DFT. The GA allows the opportunity to predict ground-state structure by just calculating energy of different off-spring structures without calculating forces on atoms or explicitly performing structure relaxations. The 2H-$MoS_2$ structure is known to be the ground state for the Mo-S system [35,42] and this structure was indeed found to be the most stable one during the GA search as shown in Fig. 4a. In addition, the ML model also identified new Mo-S configurations as stable structures. These structures were $MoS_{29}$, $MoS_{27}$, $Mo_{29}S$, and $Mo_{21}S$. A snapshot of $Mo_{21}S$ is shown in Fig. S5. We carried out similar searches for W-S and Mo-W-S system. We found that 2H-$WS_2$ structure is indeed stable [43] in ML model based convex hull plot as shown in Fig. 4b. High W and high-S containing structures ($W_{29}S$, $WS_{20}$, $W_{22}S$, $W_{28}S$ and $W_{21}S$) were also observed in W-S convex hull plot similar to Mo-S system. The stable and unstable structures are denoted with blue and red spheres respectively. The Mo-W-S convex hull diagram in Fig. 4c shows the applicability of ML and GA combined model to map energy space of multicomponent system as well. The ML based GA method is quite inexpensive due to fast formation energy ML model.

It is to be noted that classical force-fields ( such as COMB [44] and ReaxFF [45]) are also prone to finding unphysical metastable structures during GA search. Also, using the current methodology it is possible to map energy landscape of all possible multicomponent systems of 82 elements as mentioned above. For FF training, this would be unfeasible because of high-dimensional chemical combinations. After the GA with ML model, DFT calculations should be carried out only on low energy structures to reduce computational cost as an application. The ML-screened and DFT-validated structures can then be used in higher scale modeling method such as CALPHAD [46]. Most importantly, phase space mapping such as with the GA search cannot be performed with the chemical descriptors only, because it doesn’t have any insight on the crystal structure. This shows an excellent field of application for our formation-energy ML model.

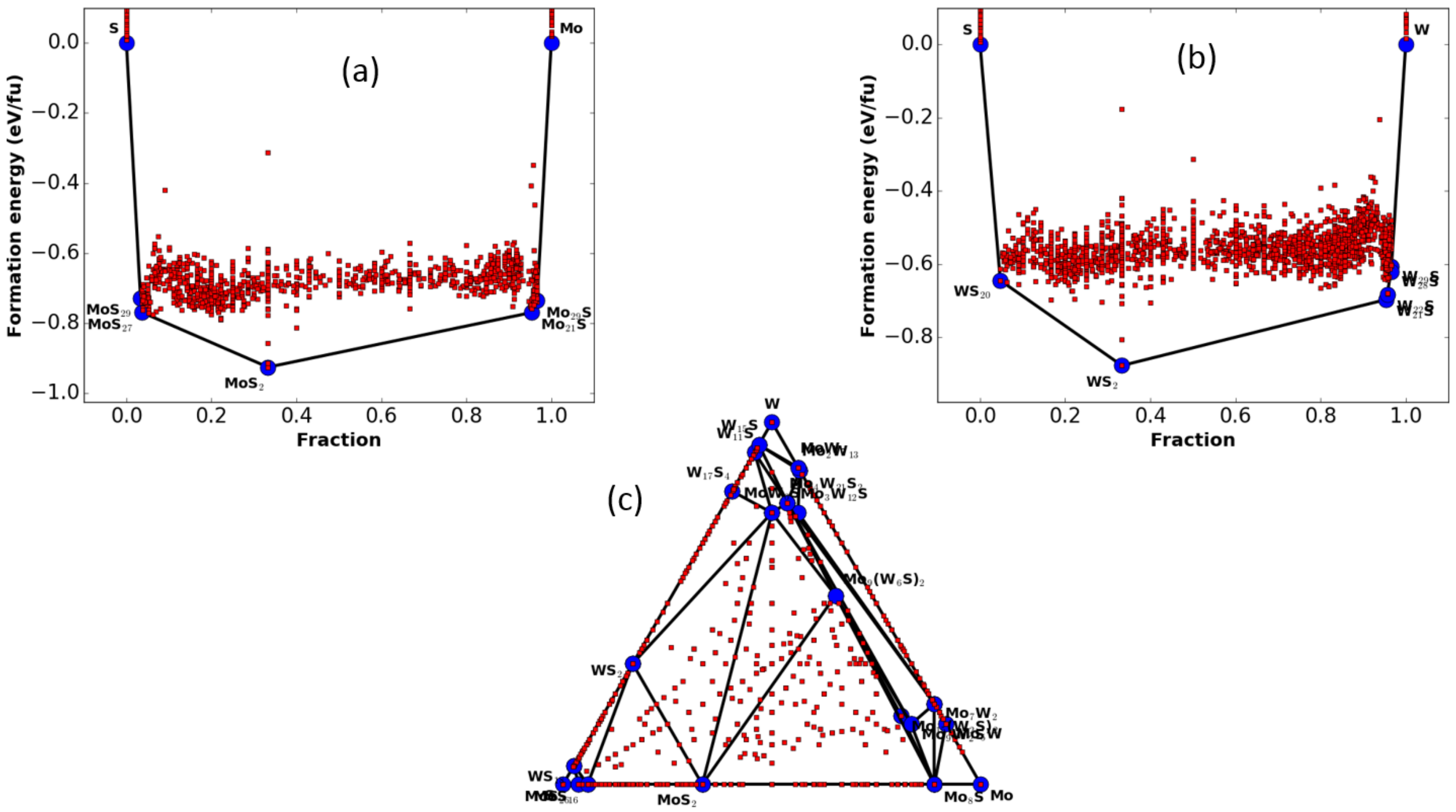

*Fig. 4) Convex-hull plot using machine learning formation energy model as energy calculator in genetic algorithm. a) Mo-S system, b) W-S system, and c) Mo-W-S system.*

## VI. CONCLUSIONS

In conclusion, we have introduced a complete set of chemo-structural descriptors and applied it to learning a wide variety of material properties, obtaining very high accuracy while training on a relatively small dataset for multicomponent systems. Although in this work the ML models were trained on specific properties, the same descriptors can be used for any other physical property as well. We have demonstrated the application of ML in materials to screen exfoliable semiconducting materials with specific requirements (like energy gap), which can drastically expedite material discovery. Integration with the evolutionary search algorithm (GA) opens a new paradigm for the accelerated investigation of high-dimensional structure and energy landscape. It also helps us understand the gap between the conventional data science and materials specific application of ML techniques. We envision that ML can be used a pre-screening tool for DFT like DFT is often used as screening tool for experiments. Genetic algorithm test for formation energy model shows some unphysical structures but those are also encountered in classical-forcefields. However, compared to intensive training process involved in conventional FFs, the present methodology should be preferred. The learnt model parameters and the computational framework are distributed publicly on the web as they can play a significant role in advancing the application of ML techniques to material science.

**Acknowledgment:**

We thank Carelyn Campbell, Kevin Garrity, Daniel Wheeler, Yuri Mishin, and Aaron Gilad Kusne at NIST for helpful discussions.

# Supplementary information: Machine learning with force-field inspired descriptors for materials: fast screening and mapping energy landscape

Kamal Choudhary, Brian DeCost, Francesca Tavazza

1 Materials Science and Engineering Division, National Institute of Standards and Technology, Gaithersburg, Maryland 20899, USA

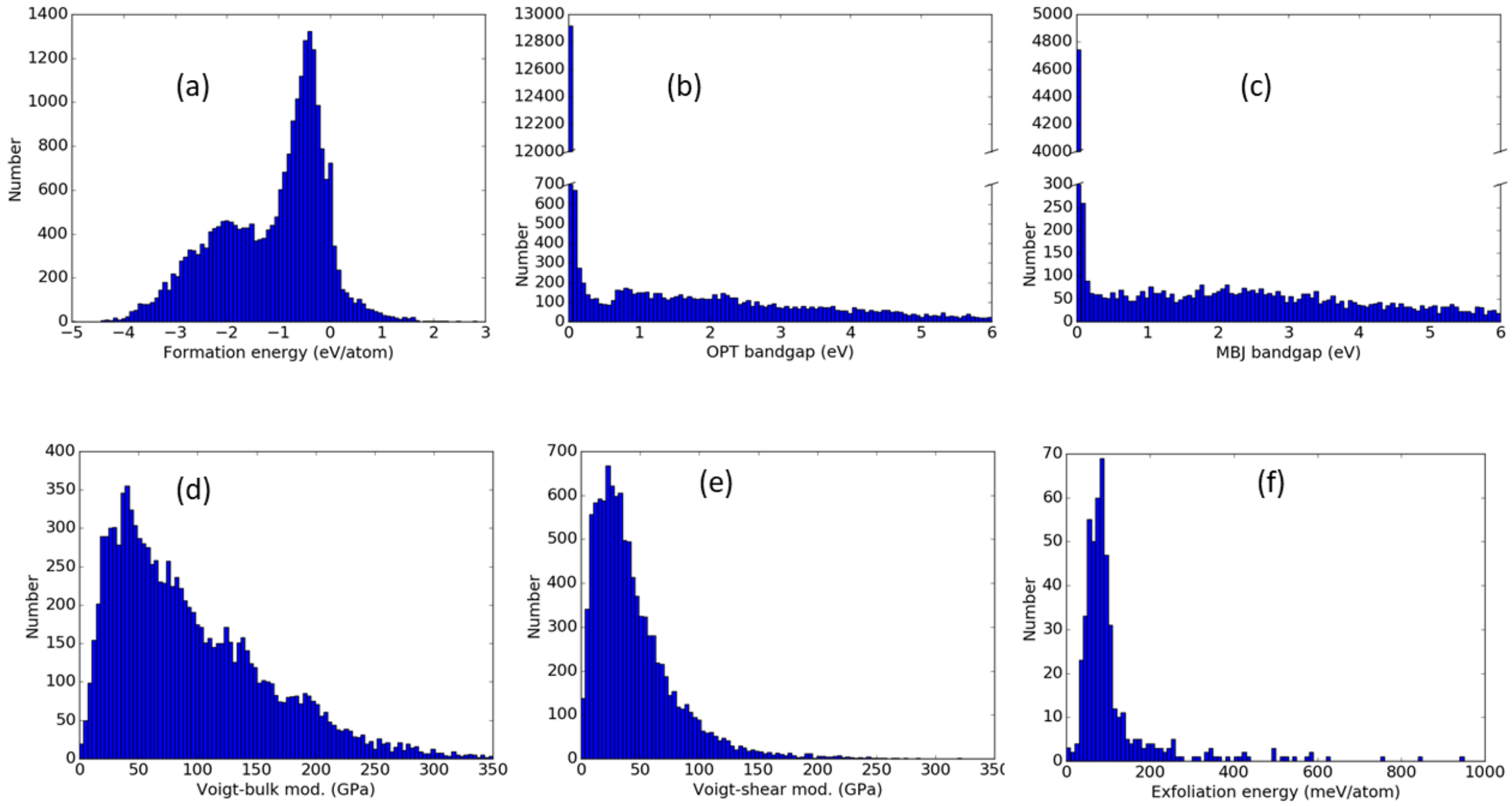

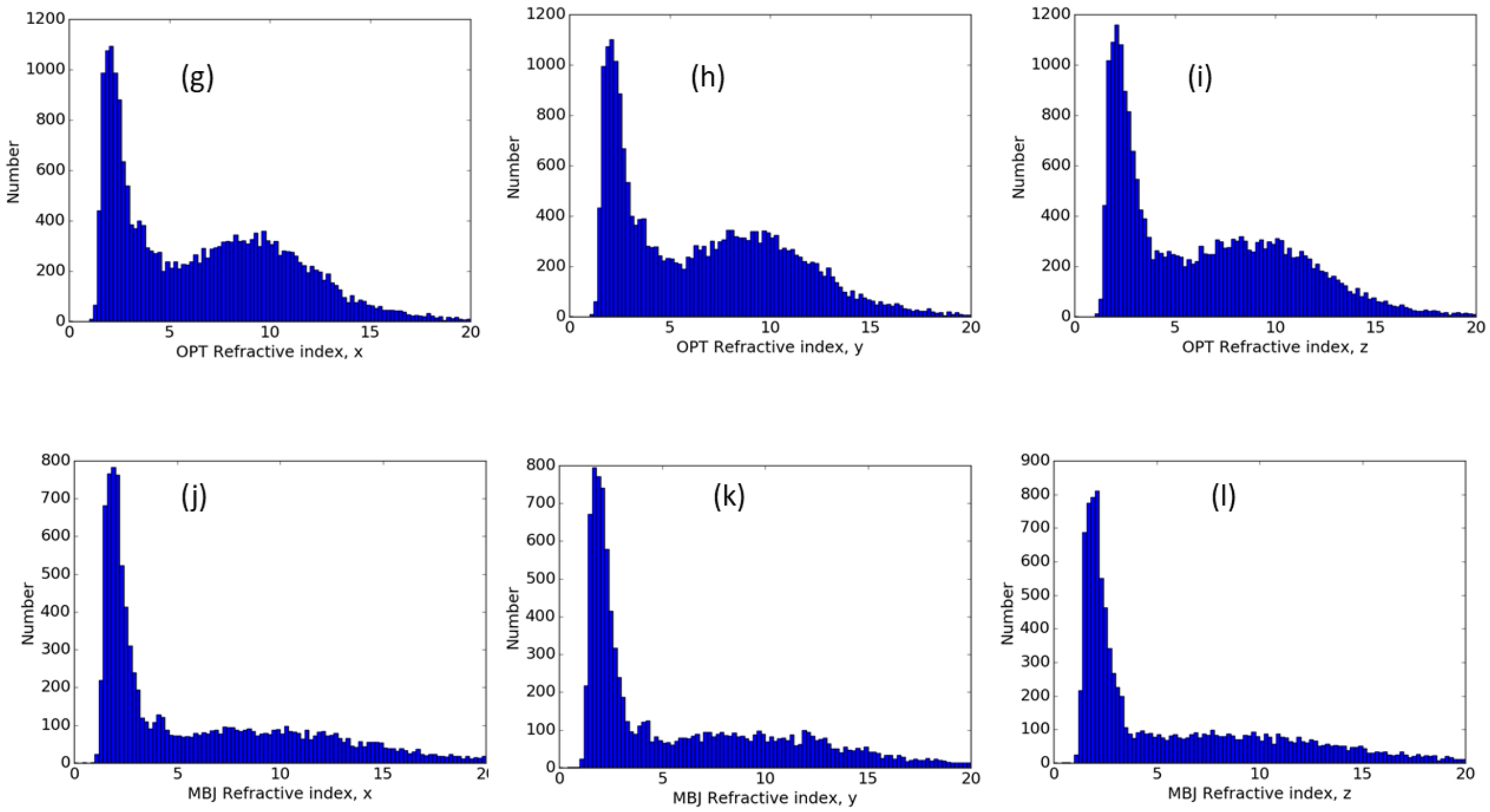

Fig. S1 Data distribution of material properties.

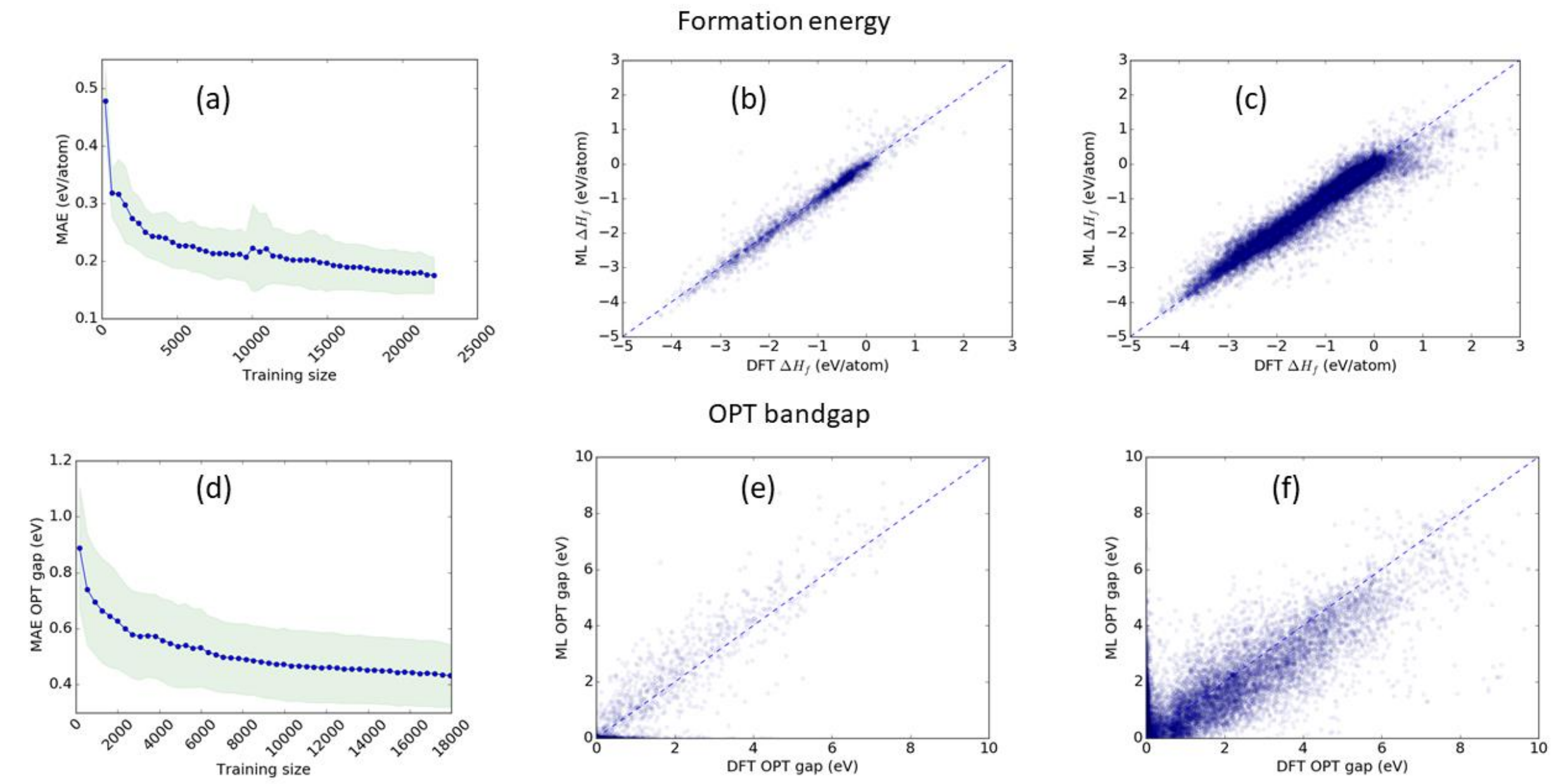

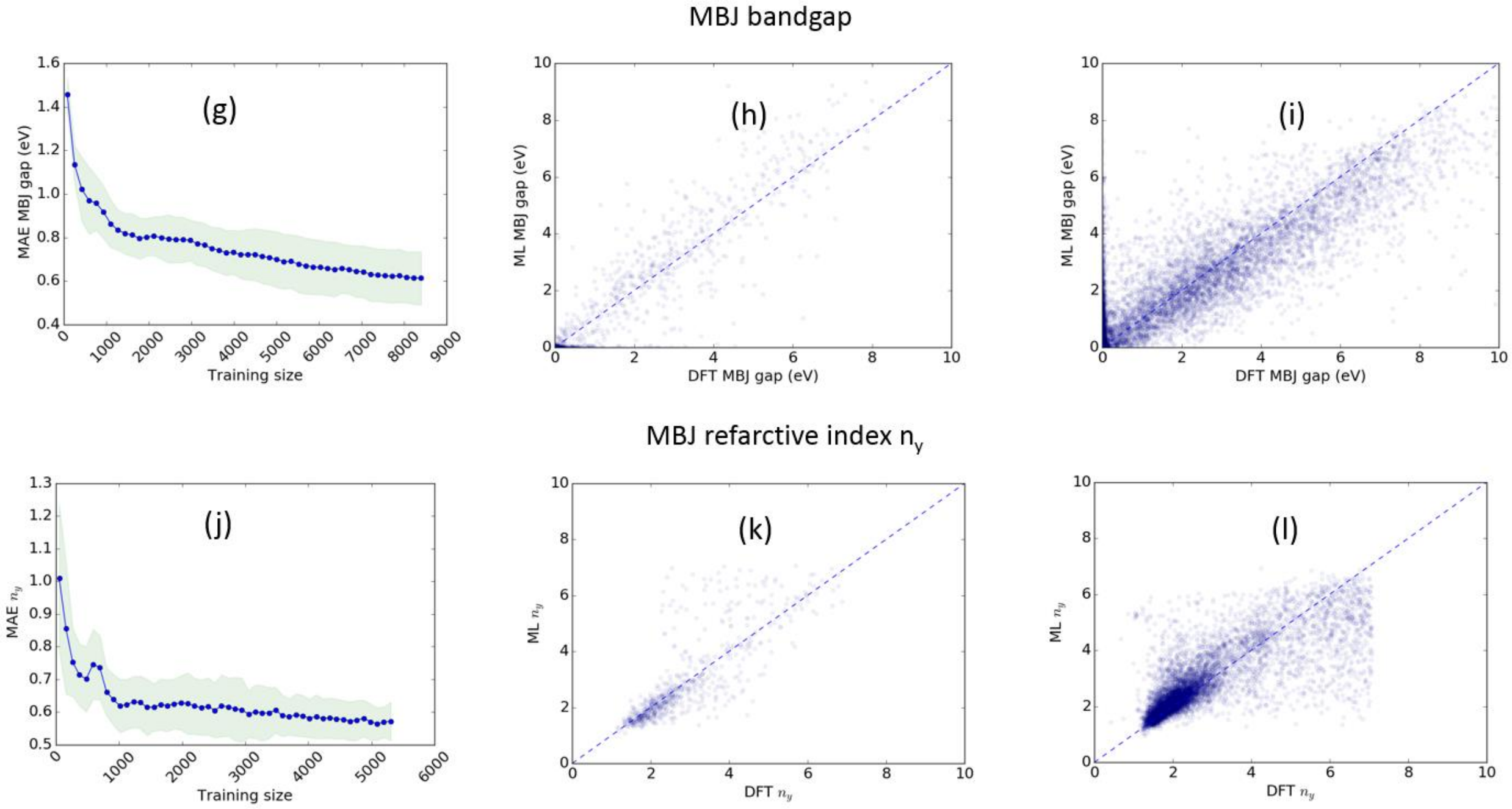
MBJ bandgap
(g)
(h)
(i)
MAE MBJ gap (eV)
Training size
ML MBJ gap (eV)
DFT MBJ gap (eV)
MBJ refarctive index n_y
(j)
(k)
(l)
MAE n_y
Training size
ML n_y
DFT n_y

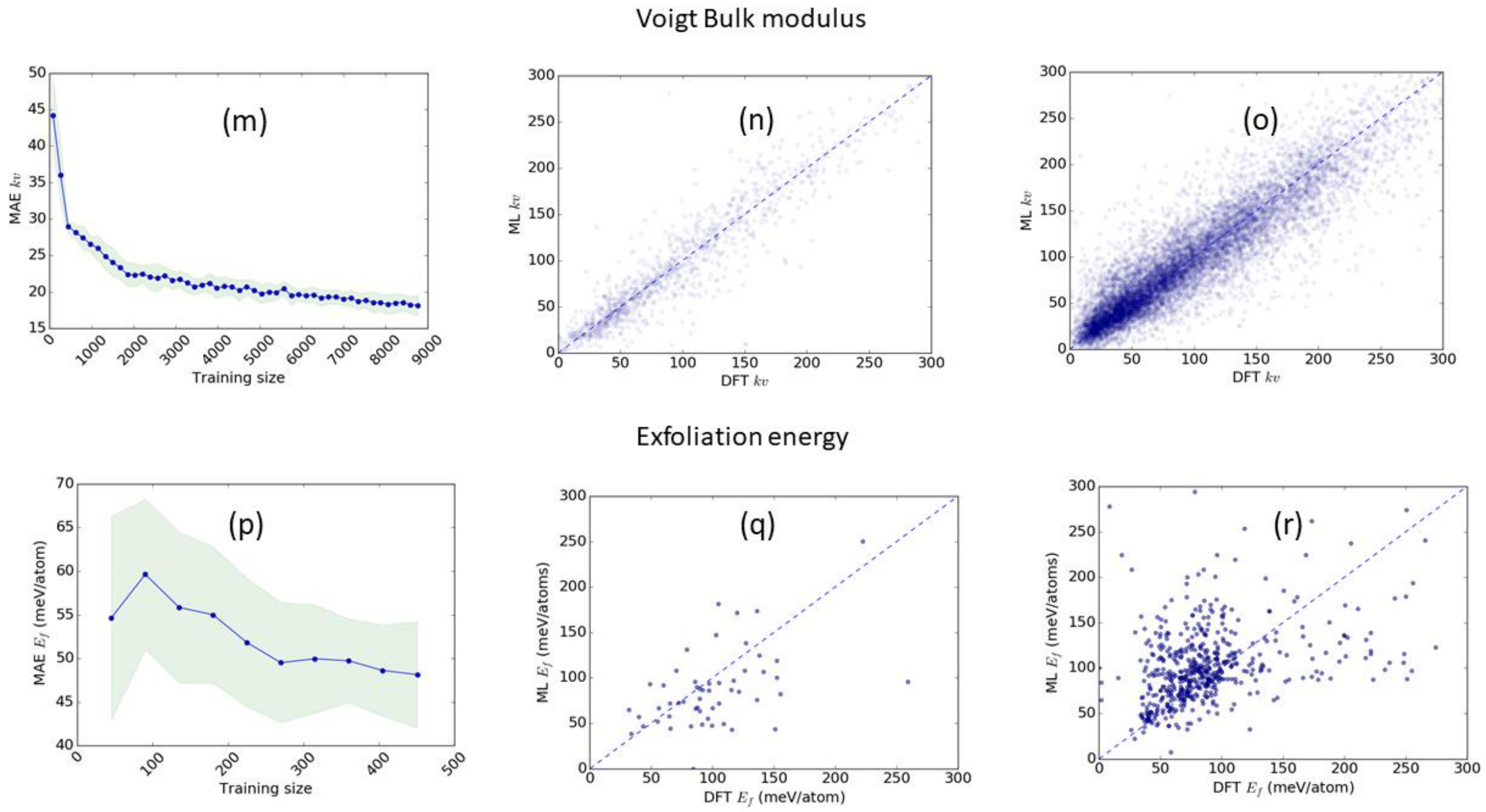

Fig. S2 Learning curve, prediction on 10 % held samples and five-fold cross-validation results of models.

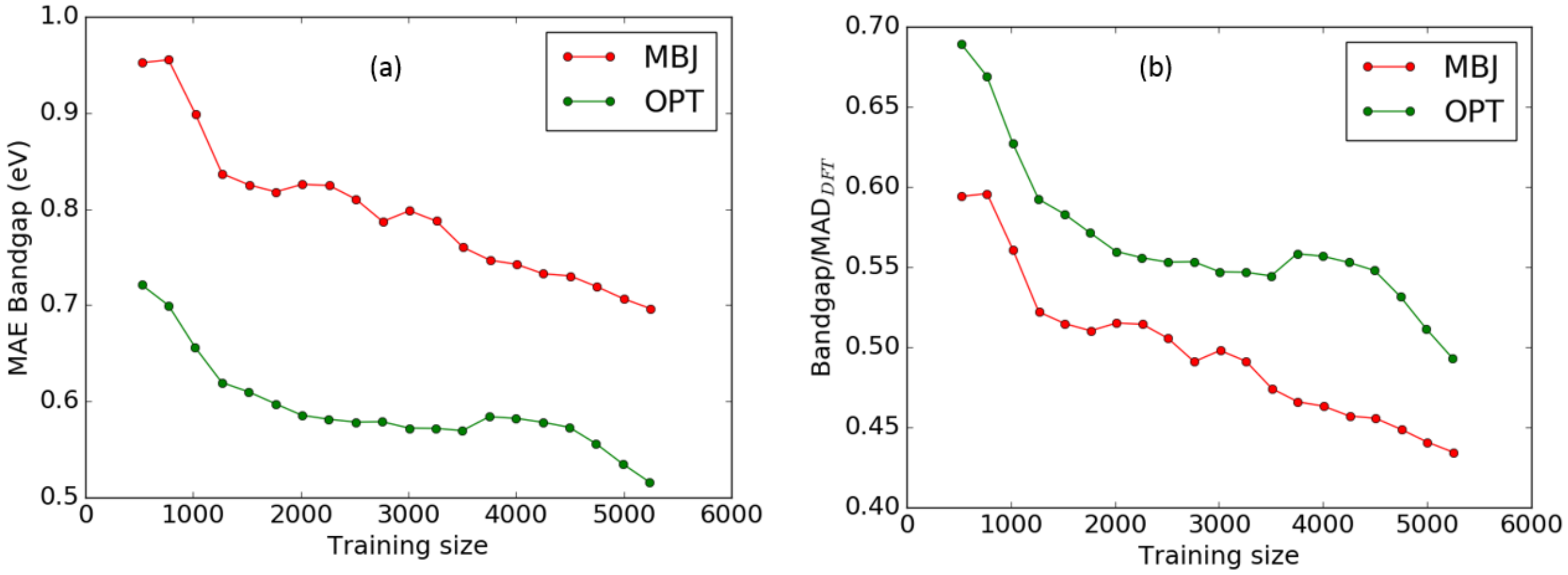

Fig S3. Comparison of learning curves for OPT and MBJ ML models.

Table S1a. Components of classical force-field inspired descriptors (CFID)

| **Descriptor name** | **Array index** | **Total number** |
|---|---|---|
| Chemical (mean_chem) | 0-437 | 438 |
| Simulation cell-size (cell) | 438-441 | 4 |
| Radial charge (mean_chg) | 442-819 | 378 |
| Radial distribution function (rdf) | 820-919 | 100 |
| Angular distribution upto first nearest neighbor cutoff (adfa) | 920-1098 | 179 |
| Angular distribution upto second nearest neighbor cutoff (adfb) | 1099-1277 | 179 |
| Dihedral distribution upto first nearest neighbor cutoff (ddf) | 1278-1456 | 179 |
| Nearest neighbor distribution (nn) | 1457-1556 | 100 |
| **Total** | | 1557 |

Table S1b. Details of element based chemical descriptors

| **Descriptor name** | **Details** |
|---|---|
| jv_enp | Energy per atom of an element from JARVIS-DFT |
| KV | Bulk modulus of an element from JARVIS-DFT |
| GV | Shear modulus of an element from JARVIS-DFT |
| C-m (m=0 to 35) | Elastic constants of an element from JARVIS-DFT (total 36) |
| op_eg | OptB88vdW bandgap during SCF for an element |
| mop_eg | OptB88vdW bandgap during linear optics for an element |
| voro_coord | Voronoi coordination number of an elemental-crystal structure |
| ndunfilled | Number of unfilled d-orbitals |
| ndvalence | Number of valence d-orbitals |
| nsunfilled | Number of unfilled s-orbitals |
| nsunfilled | Number of valence s-orbitals |
| npunfilled | Number of unfilled p-orbitals |
| npvalence | Number of valence p-orbitals |
| nfunfilled | Number of unfilled f-orbitals |
| nfvalence | Number of valence f-orbitals |
| first_ion | First ionization energy of an element |
| oq_bg | OQMD bandgap for an element |
| elec_aff | Electron affinity |
| vol_pa | Volume per atom of an element |
| hfus | Heat of fusion of an element |
| oq_enp | OQMD energy per atom |
| Polariz | Polarizability |
| Z | Atomic number |
| X | Electronegativity |
| row | Row number in the periodic table |
| column | Column number in the periodic table |
| max_oxid_s | Maximum oxidation state |
| min_oxid_s | Minimum oxidation state |
| block | s,p,d,f block assigned to 0,1,2,3 blocks |
| is_alkali | Is it alkali element 0/1 |
| is_alkaline | Is it alkaline element 0/1 |
| is_metalloid | Is it metalloid element 0/1 |
| is_noble_gas | Is it noble gas element 0/1 |
| is_transition_metal | Is it transition element 0/1 |
| is_metalloid | Is it metalloid element 0/1 |
| is_halogen | Is it halogen element 0/1 |
| is_lanthanoid | Is it lanthanoid element 0/1 |
| is_actinoid | Is it actinoid element 0/1 |
| atom_mass | Atomic mass |
| atom_rad | Atomic radii |
| therm_cond | Thermal conductivity |
| mol_vol | Molar volume |
| bp | Boiling point |
| mp | Melting point |
| avg_ion_rad | Average ionic radii |
| polzbl | Polarizability |
| e1 | Static dielectric function in x-direction from JARVIS-DFT using OptB88vdW functional |
| e2 | Static dielectric function in y-direction from JARVIS-DFT using OptB88vdW functional |
| e3 | Static dielectric function in z-direction from JARVIS-DFT using OptB88vdW functional |
| me1 | Static dielectric function in x-direction from JARVIS-DFT using TB-mBJ potential |
| me2 | Static dielectric function in y-direction from JARVIS-DFT using TB-mBJ potential |
| me3 | Static dielectric function in z-direction from JARVIS-DFT using TB-mBJ potential |

Addition (‘add’), multiplications (‘mult’), subtraction (‘subs’) and quotient (‘divi’) of hfus, polzbl, first_ion_en, mol_vol, bp,mp, mol_vol, mol_vol, therm_cond and voro_coord were performed to give additional chemical descriptors.

Table S1c. Details of simulation cell-size based descriptors

| Descriptor name | Details |
|---|---|
| cell_0 | Volume per atom of the cell |
| cell_1 | Logarithm of volume per atom of the cell |
| cell_2 | Packing fraction |
| cell_3 | Density |

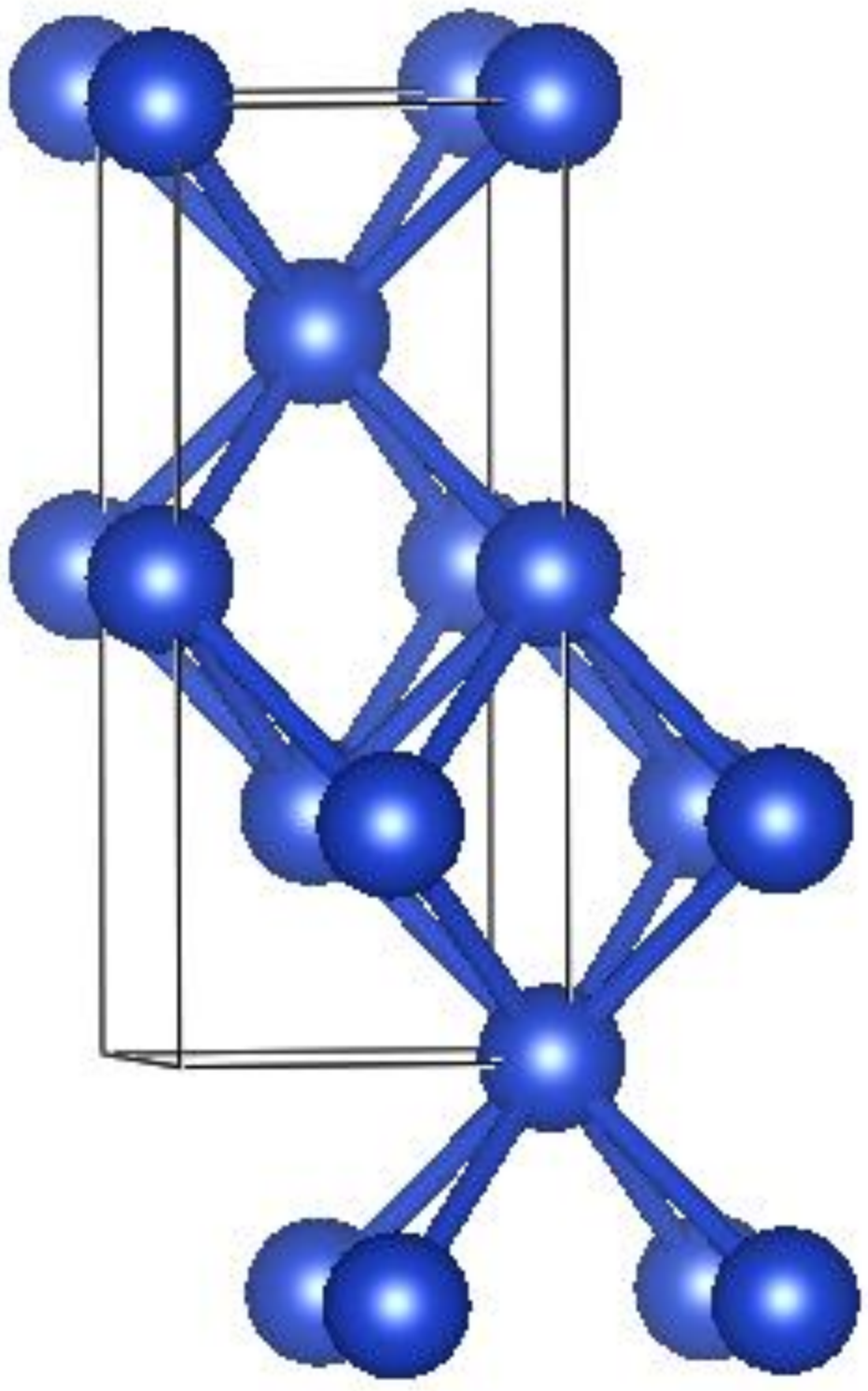

Fig. S4 Atomic structure of Cu (I4/mmm) obtained during GA search of ML assisted GA training of Cu.

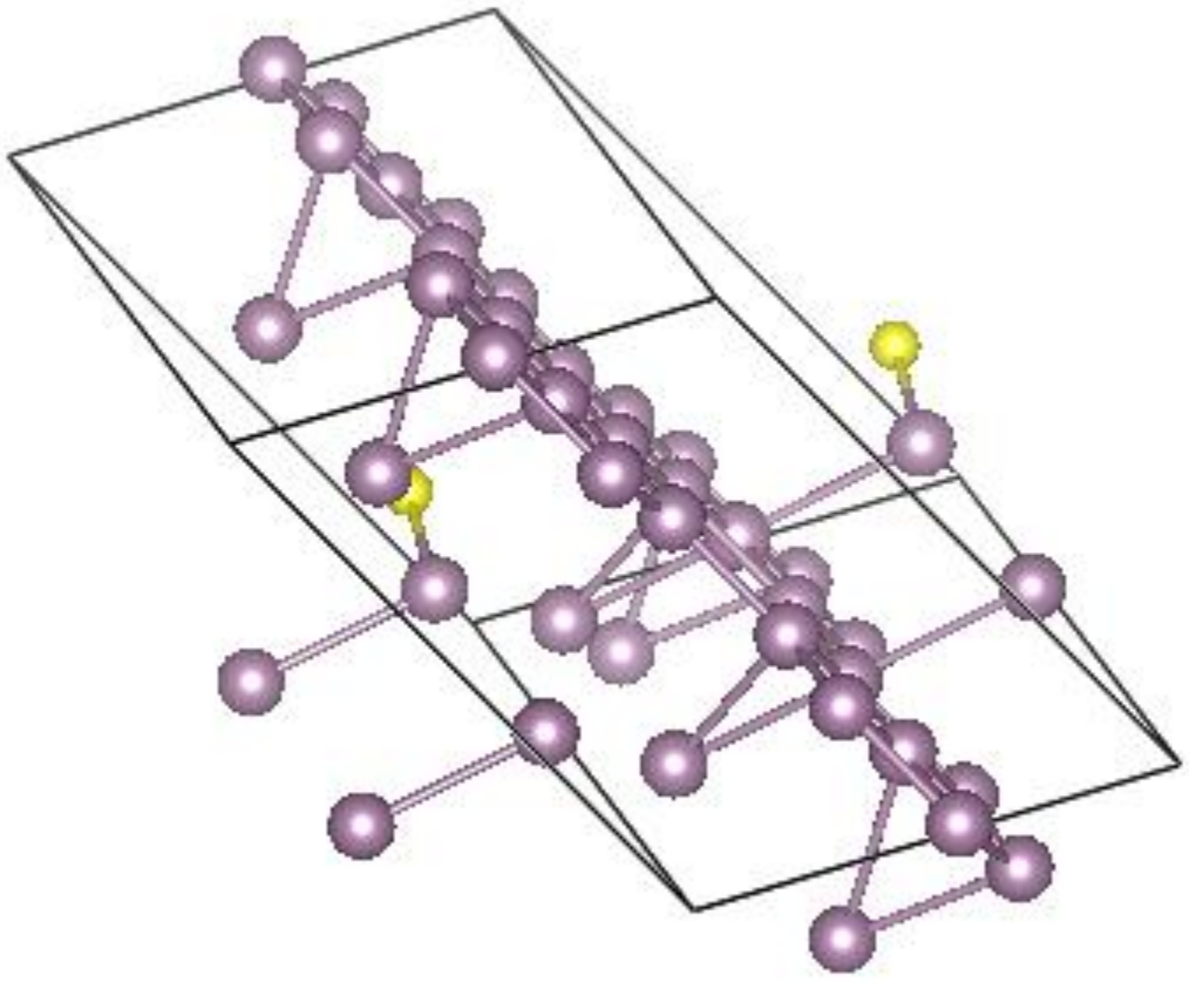

Fig. S5 Atomic structure of $Mo_{21}S$ obtained during GA search of ML assisted GA training of Mo-S system.